\documentclass[12pt]{article}
\usepackage[export]{adjustbox}
\usepackage{graphicx}
\usepackage{graphics}
\usepackage{epsfig}
\usepackage{amssymb}
\usepackage{amsmath}
\usepackage{color}
\usepackage{textcomp}
\usepackage{latexsym}
\usepackage{physics}

\usepackage[margin=2.5cm,footskip=0.6cm]{geometry}
\begin{document}

\begin{center}
{\Large{\bf Velocity Distribution and Diffusion of an Athermal Inertial Run-and-Tumble Particle in a Shear-Thinning Medium}} \\
\ \\
\ \\
by \\
Sayantan Mondal and Prasenjit Das\footnote{prasenjit.das@iisermohali.ac.in} \\
Department of Physical Sciences, Indian Institute of Science Education and Research -- Mohali, Knowledge City, Sector 81, SAS Nagar 140306, Punjab, India. \\
\end{center}

\begin{abstract}
\noindent We study the dynamics of an athermal inertial active particle moving in a shear-thinning medium in $d=1$. The viscosity of the medium is modeled using a Coulomb-tanh function, while the activity is represented by an asymmetric dichotomous noise with strengths $-\Delta$ and $\mu\Delta$, transitioning between these states at a rate $\lambda$. Starting from the Fokker-Planck~(FP) equation for the time-dependent probability distributions $P(v,-\Delta,t)$ and $P(v,\mu\Delta,t)$ of the particle's velocity $v$ at time $t$, moving under the influence of active forces $-\Delta$ and $\mu\Delta$ respectively, we analytically derive the steady-state velocity distribution function $P_s(v)$, explicitly dependent on $\mu$. Also, we obtain a quadrature expression for the effective diffusion coefficient $D_e$ for the symmetric active force case~($\mu=1$). For a given $\Delta$ and $\mu$, we show that $P_s(v)$ exhibits multiple transitions as $\lambda$ is varied. Subsequently, we numerically compute $P_s(v)$, the mean-squared velocity $\langle v^2\rangle(t)$, and the diffusion coefficient $D_e$ by solving the particle's equation of motion, all of which show excellent agreement with the analytical results in the steady-state. Finally, we examine the universal nature of the transitions in $P_s(v)$ by considering an alternative functional form of medium's viscosity that also capture the shear-thinning behavior.
\end{abstract}

\newpage
\section{Introduction}

\label{sec1}
The physics of active particles has blossomed into a vibrant field of research in recent years due to its wide-ranging applications in diverse scientific disciplines~\cite{S10,MJSTJR13,J24}.  Active matter systems represent a unique class of non-equilibrium systems, inherently maintained out of equilibrium by the internal forces of their constituents~\cite{CPDLG20,JGMJC22,KGS23}. These constituents exhibit self-propulsion by harvesting energy from their surroundings and converting it into sustained, directed persistent motion. In doing so, they break the detailed balance condition at the microscopic level, leading to a fascinating array of phenomena observable both at the single-particle scale and in collective dynamics.

Over the past few decades, Active Brownian Particles (ABPs) have emerged as a widely studied minimal model for capturing self-propulsion in active matter systems~\cite{PMWBL12,AH12,USAG18}. ABPs provide a simple yet effective description of self-driven particles that persistently move in a given direction while subject to both translational and rotational diffusion. Example of such systems include self-propelled colloids, motile cells, bacteria, algae, and synthetic microswimmers such as Janus particles~\cite{CRHCG16,AKTP21,YJRB23,AJAYPJ23,SP24}. The theoretical model of ABPs combines overdamped dynamics with a self-propulsion force to describe their intrinsic behavior. The self-propulsion force is inherently stochastic, originating from interactions with the surroundings, and is often modeled as Gaussian white noise.~\cite{IGFGC12,IJFHCT13,AJRMYMJ13,JRG15,AH23}. 

Recently, an alternative framework for modeling active particles has been proposed. In this framework, the self-propulsion force is characterized by a stochastic Ornstein-Uhlenbeck process, which captures persistent motion through exponentially correlated noise~\cite{GE36,YM12,ECMJPF16}. These particles are known as Active Ornstein-Uhlenbeck Particles (AOUPs). The AOUP model captures key characteristics of active systems such as persistent motion, wetting of particles near an obstacle or boundary, and notably, motility-induced phase separation (MIPS)~\cite{MJ15,LUAA19,DJMCJF21,ERM22,LDACH24}. AOUPs have also been studied as a model for passive tracer dynamics in an active bacterial bath, as well as for investigating the collective dynamics of cells~\cite{CMNALD14,CMLR17}. Notably, the AOUP framework offers the advantage of providing exact or approximate analytical solutions for a wide range of problems, making it a powerful tool for modeling self-propelled motion~\cite{TPJ15,ARJ17}.

Previous studies on ABPs and AOUPs primarily focused on the overdamped dynamics of microscopic active particles, effectively neglecting inertial effects. However, inertial effects become non-negligible for macroscopic active particles moving in gaseous or low-viscosity media, posing new challenges for their theoretical modeling~\cite{H20}. Macroscopic active particles are abundant in nature, with examples including birds, insects, and fish~\cite{AALITASL0EM14,JRUMTHV20}. Beyond living systems, they also appear in inanimate experimental setups, such as driven granular materials, autonomous hexbots, and minibots~\cite{MG13,MFJG18,0V19}. These experimental realizations have further inspired extensive theoretical studies on the role of inertia in active matter systems~\cite{GRH22,LASHR22,LRH24}.
 
The dynamics of a single particle subjected to linear friction and Gaussian white noise is a well-established paradigm in nonequilibrium statistical physics, commonly used to illustrate fundamental principles of stochastic processes in dissipative systems~\cite{C85,H96,V08}. Linear friction is the characteristic of wet systems, such as simple liquids and colloidal suspensions, where the frictional force $F_f(v)$ is typically proportional to the velocity, i.e., $F_f(v)\propto v$. This force originates from interactions with the surrounding bath particles and aligns with Stokes' law. Again, to describe the dynamics of stochastically driven particles sliding on solid surfaces, Coulomb friction—also known as solid or dry friction—is widely employed~\cite{LVINCI,C21,P05,H05,AN11,PSM17}. Unlike linear friction, the Coulomb friction force exhibits an abrupt transition from zero to a finite value and maintains a constant magnitude that always opposes the particle's direction of motion, $\hat{n}$, i.e., $F_f(v)\propto -\hat{n}$. It is essential for understanding phenomena such as stick-slip motion, the dynamics of dense granular fluids, the behavior of rotators in granular media, granular particle flow, and the ratchet motion of solid objects on vibrating surfaces~\cite{RDAL98,YM98,PSM16,PSM18}. Aside from problems involving Gaussian white noise, there are studies that explore Coulomb friction with non-Gaussian noise. For example, Geffert \textit{et al.} studied the impact of colored noise on a pure Coulomb friction model using the Unified Colored Noise Approximation (UCNA)~\cite{PP87,PP95,PW17}. Their work revealed a stick-slip transition linked to a critical threshold in the noise correlation time, highlighting the role of temporal correlations in driving this behavior. Recently, Antonov \textit{et al.} conducted experimental and numerical studies on the dynamics of inertial active particles under Coulomb friction, observing excellent agreement between their experimental and numerical results~\cite{ALACH24}. These studies focused on key quantities of interest, including the position and velocity distribution functions, the velocity autocorrelation function, the mean-squared displacement, diffusion coefficients, etc.
 
A variation of the Coulomb friction model that eliminates the abrupt jump in velocity is known as Coulomb-tanh friction model~\cite{SAS07,EVPP16,CSA21}. In this model, the friction force $F_f(v)$ is described by a hyperbolic tangent function: $F_f(v)\propto \tanh(g v)$, where $g$ is a parameter that governs the smoothness of the transition from viscous to Coulomb friction. The dynamics under this frictional force partially mimic the behavior of a particle in a shear-thinning environment, e.g., polymer solutions, biological fluids, paints, etc~\cite{J09}. The linear friction and Coulomb friction models described earlier can be obtained from Coulomb-tanh friction in the limits $v\rightarrow 0$ and $v\rightarrow\infty$, respectively. Other models, such as the Coulomb-viscous, Stribeck friction, and Tustin emperical models, also capture shear-thinning behavior~\cite{ALACH24,EVPP16}. However, the functions describing these models are not as smooth as the Coulomb-tanh friction model, making them non-differentiable at certain points. Additionally, they involve multiple parameters, adding to their complexity.

In a recent paper, Lequy \textit{et al.} analyzed the stochastic dynamics of particles under Coulomb-tanh friction by deriving the velocity distribution through analytical methods, revealing how nonlinear friction influences steady-state dynamics~\cite{TA23}. Most recently, Dutta \textit{et al.} studied the non-equilibrium stationary state dynamics of a one-dimensional inertial run-and-tumble particle, both in the presence and absence of harmonic trapping~\cite{DASU24,DAU24}. They found that the system's dynamical behavior, characterized by the position distribution, velocity distribution, position-velocity joint probability distribution, mean-squared displacement, etc., depends on two intrinsic time scales: the inertial time scale and the active time scale. However, the impact of activity on athermal inertial particles moving in shear-thinning environments remains largely unexplored. In this paper, we investigate the dynamics of an inertial run-and-tumble particle subjected to Coulomb–tanh friction and driven by an asymmetric dichotomous noise in $d=1$. This noise captures the stochastic switching dynamics of particle's velocity, as a particle moves under a constant force in either direction and randomly switches states at a given rate. To the best of our knowledge, this is the first study of its kind. 

Given this background, the structure of the paper is as follows: In Sec.~\ref{sec2}, we provide the details of the model and present the analytical calculations for the steady-state velocity distribution function, $P_s(v)$ and the effective diffusion coefficient, $D_e$. The numerical validation of our analytical results is discussed in Sec.~\ref{sec3}. In Sec.~\ref{sec4}, we test the universal nature of the transitions in $P_s(v)$. Finally, we summarize our results in Sec.~\ref{sec5}.

\section{Model and Analytical Results}\label{sec2}
We consider the motion of an athermal inertial active particle moving in a shear thinning medium, where viscosity of the medium is modeled by Coulomb-tanh friction. The dynamics of the particle is described by Langevin’s equation of motion as
\begin{eqnarray}
\label{apeqn1}
m \frac{d\vec{v}}{dt} = -\gamma \tanh\left(\frac{|\vec{v}|}{v_0}\right) \hat{n} + \vec{\zeta}.
\end{eqnarray}
In Eq.~(\ref{apeqn1}), $m$ and $\vec{v}$ represent the mass and velocity of the particle, respectively. $\hat{n}$ is the unit vector along the direction of $\vec{v}$. The first term on the right-hand side, $-\gamma \tanh\left(|\vec v|/v_0\right)$, represents the Coulomb-tanh friction force, where $\gamma$ is the friction strength, and $v_0$ sets a characteristic velocity scale determined by the properties of the medium. The term $\vec{\zeta}$ represents the active force, which is modeled by a dichotomous Markov process~\cite{WR83,J84,I06}.

For simplicity, we solve the Eq.~(\ref{apeqn1}) in $d=1$. In this case, the equation of motion reduces to
\begin{eqnarray}
\label{apeqn1a}
m \frac{dv}{dt} = -\gamma\tanh\left(\frac{v}{v_0}\right) + \zeta.
\end{eqnarray}
Here, we assume that $\zeta$ can take only two possible values, $\mu\Tilde{\Delta}$ and $-\Tilde{\Delta}$, with equal probability, transitioning between them with a probability of $\Gamma dt$ in a time interval $dt$, where $\Gamma$ denotes the flipping rate. Here, $\mu$ is a positive parameter that quantifies the asymmetry in the active force. The mean value and autocorrelation of the active force are given by
\begin{eqnarray}
\label{apeqn2}
\langle \zeta(t)  \rangle =\frac{(\mu-1)\Tilde{\Delta}}{2}\quad\text{and}\quad 
\langle \zeta(t) \zeta(t^{\prime}) \rangle = \frac{\left(\mu+1\right)^2\Tilde{\Delta}^2}{4} \exp\left(-2 \Gamma |t - t^{\prime}|\right).
\end{eqnarray}
This asymmetric dichotomous Markov noise~(DMN) effectively mimics the run-and-tumble motion of living organisms in $d=1$. Here, we emphasize that the active force parameters are independent of the medium's viscosity since activity is an intrinsic property of the particle.

Next, we nondimensionalize Eq.~(\ref{apeqn1a}) by rescaling time and velocity as
\begin{eqnarray}
\label{apeqn2a}
t = \frac{mv_0}{\gamma}t^\prime \quad \text{and} \quad v = v_0v^\prime,
\end{eqnarray}
where $t^\prime$ and $v^\prime$ are the dimensionless time and velocity, respectively. Using these substitutions, the nondimensional form of Eq.~(\ref{apeqn1a}) becomes (dropping the primes)
\begin{eqnarray}
\label{apeqn1b}
\frac{dv}{dt} = -\tanh\left(v\right) + \xi,
\end{eqnarray}
The scaled asymmetric active force $\xi(t)$ satisfies the following properties:
\begin{eqnarray}
\label{apeqn2b}
\langle \xi(t) \rangle = \frac{(\mu-1)\Delta}{2}  \quad \text{and} \quad \langle \xi(t) \xi(t^{\prime}) \rangle = \frac{\left(\mu+1\right)^2\Delta^2}{4}\exp\left(-2\lambda|t - t^{\prime}|\right),
\end{eqnarray}
with $\Delta = \Tilde{\Delta}/\gamma$ and $\lambda=mv_0\Gamma/\gamma$ being the scaled active force strength and flipping rate of the active force, respectively.

We determine the stable fixed points of the particle's dynamics, which yield two maximum possible velocity of the particle, $v_a$ and $v_b$, corresponding to the active force states $\mu\Delta$ and $-\Delta$, respectively. At these fixed points, we find:
\begin{eqnarray}
\label{apeqn3}
\tanh{\left(v_a\right)} = \mu\Delta \quad \text{and} \quad \tanh\left(v_b\right) = -\Delta,
\end{eqnarray}
which implies,
\begin{eqnarray}
\label{apeqn4}
v_a = \tanh^{-1}(\mu\Delta) \quad \text{and} \quad v_b = -\tanh^{-1}\Delta.
\end{eqnarray}
This result implies that the particle's velocity always lies within the range $v_b=-\tanh^{-1}\Delta$ to $v_a=\tanh^{-1} (\mu\Delta)$. Clearly, the quantities $\Delta$ and $\mu\Delta$ must be less than unity, with $\mu$ being a positive number.

\subsection{Steady State Velocity Distribution Function $P_s(v)$}\label{2a}
We define $P(v,\mu\Delta, t)$ as the probability density of the particle having velocity $v$ at time $t$, given the active force $\mu\Delta$. Similarly, $P(v,-\Delta, t)$ represents the probability density of the particle having velocity $v$ at time $t$, given the active force $-\Delta$. The corresponding Fokker-Planck~(FP) equations for these probability densities, derived from the dynamical equation~(\ref{apeqn1b}), are as follows
\begin{eqnarray}
\label{apeqn5}
&&\frac{\partial}{\partial t} P(v,\mu\Delta, t) = -\frac{\partial}{\partial v} \left[ -\tanh{\left(v\right)}+\mu\Delta  \right]P(v,\mu\Delta,t) - \lambda\left[P(v,\mu\Delta,t) - P(v,-\Delta,t)\right], \\
\label{apeqn6}
&&\frac{\partial}{\partial t} P(v,-\Delta, t) = -\frac{\partial}{\partial v} \left[ -\tanh{\left(v\right)}-\Delta  \right]P(v,-\Delta,t) + \lambda\left[P(v,\mu\Delta,t) - P(v,-\Delta,t)\right].
\end{eqnarray}
In Eqs.~(\ref{apeqn5}) and (\ref{apeqn6}), the first term on RHS represents the deterministic evolution of the probability densities under the influence of friction force $-\tanh(v)$ and active force $\mu\Delta$ or $-\Delta$, and redistributing them in velocity space. The second term accounts for the stochastic switching of the active force, which redistributes the probability between the two force states at a fixed velocity $v$. Here, we are interested in the probability density $P(v,t)$, which represents the probability of the particle having $v$ at time $t$, irrespective of the state of active force on it. It is given by $P(v,t)=P(v,\mu\Delta,t)+P(v,-\Delta,t)$. Additionally, we define a new function $Q(v,t)$ as $Q(v,t)=\lambda[P(v,\mu\Delta,t)-P(v,-\Delta,t)]$.

Using Eqs.~(\ref{apeqn5}) and (\ref{apeqn6}), the dynamical equations for $P(v,t)$ and $Q(v,t)$ are given as
\begin{eqnarray}
\label{apeqn7}
&&\frac{\partial}{\partial t} P(v,t) = -\frac{\partial}{\partial v} \left[ -\tanh{\left(v\right)}+\frac{(\mu-1)\Delta}{2}\right]P(v,t) - \frac{(\mu+1)\Delta}{2\lambda}\frac{\partial}{\partial v} Q(v,t) , \\
\label{apeqn8}
&&\frac{\partial}{\partial t} Q(v,t) = -\frac{\partial}{\partial v} \left[ -\tanh{\left(v\right)}+\frac{(\mu-1)\Delta}{2}\right]Q(v,t) - 2\lambda Q(v,t) - \frac{(\mu+1)\Delta\lambda}{2}\frac{\partial}{\partial v}P(v,t).
\end{eqnarray}
To solve Eq.~(\ref{apeqn8}), we impose a natural assumption on the initial condition: in the infinite past, the velocity $v$ and the active force $\xi$ were statistically independent. This implies that $Q(v, t)$  satisfies $Q(v, -\infty) = 0$. Under this condition, the solution to Eq.~(\ref{apeqn8}) is given by:
\begin{eqnarray}
\label{apeqn9}
Q(v,t)=-\int^{t}_{-\infty}dt^{\prime} \exp\left[-\left\{2\lambda-\frac{\partial}{\partial v} \left( \tanh{\left(v\right)}-\frac{(\mu-1)\Delta}{2}\right)  \right\}(t - t^{\prime})\right] \frac{(\mu+1)\Delta\lambda}{2}
\frac{\partial}{\partial v} P(v, t^{\prime}).
\end{eqnarray}
Substituting the expression for $Q(v,t)$ in Eq.~(\ref{apeqn7}), we obtain a closed evaluation equation for $P(v,t)$ as follows:
\begin{eqnarray}
\label{apeqn10}
\frac{\partial}{\partial t} P(v, t) &=& -\frac{\partial}{\partial v} \left[ -\tanh{\left( v\right)+\frac{(\mu-1)\Delta}{2}}  \right]P(v,t) \nonumber \\
&& +\frac{(\mu+1)^2\Delta^2}{4} \frac{\partial}{\partial v} \int^{t}_{-\infty} dt^{\prime} \exp\left[-\left\{2\lambda-\frac{\partial}{\partial v} \left(\tanh{\left(v\right)}-\frac{(\mu-1)\Delta}{2}\right)  \right\}(t - t^{\prime})\right]
\frac{\partial}{\partial v} P(v, t^{\prime}).
\end{eqnarray}
It is not possible to solve Eq.~(\ref{apeqn10}) to obtain the complete time evolution of $P(v,t)$, as it involves derivatives of $P(v,t^{\prime})$ of infinitely high order. However, with natural boundary conditions, we can determine the stationary solution $P_s(v)$.

To calculate $P_s(v)$, we begin with the steady-state forms of Eqs.~(\ref{apeqn7}) and (\ref{apeqn8}):
\begin{eqnarray}
\label{apeqn11}
&& 0 = -\frac{\partial}{\partial v} \left[ -\tanh{\left(v\right)} +\frac{(\mu-1)\Delta}{2}\right]P_s(v) - \frac{(\mu+1)\Delta}{2\lambda}\frac{\partial}{\partial v} Q_s(v) , \\
\label{apeqn12}
&& 0 = -\frac{\partial}{\partial v} \left[ -\tanh{\left( v\right)}+\frac{(\mu-1)\Delta}{2}+2\lambda\right]Q_s(v)  - \frac{(\mu+1)\Delta\lambda}{2}\frac{\partial}{\partial v}P_s(v).
\end{eqnarray}
Equation~(\ref{apeqn11}) provides
\begin{eqnarray}
\label{apeqn13}
\left[ -\tanh{\left(v\right)+\frac{(\mu-1)\Delta}{2}} \right]P_s(v) + \frac{(\mu+1)\Delta}{2\lambda} Q_s(v)=C.
\end{eqnarray}
Since, the dynamical equation [Eq.~(\ref{apeqn1b})] for $v$ is deterministically stable, we can conclude that $C=0$. So the Eq.~(\ref{apeqn13}) becomes
\begin{eqnarray}
\label{apeqn14}
 Q_s(v)=\frac{2\lambda}{(\mu+1)\Delta}\left[\tanh{\left(v\right)-\frac{(\mu-1)\Delta}{2}} \right]P_s(v).
\end{eqnarray}
 We insert the expression for $Q_s(v)$ and $Q_s^{\prime}(v)$ in Eq.~(\ref{apeqn12}) to obtain
\begin{eqnarray}
\label{apeqn16}
P^{\prime}_s(v)&=&\left\{\frac{ [\sech^2(v)-\lambda][2\tanh(v) - (\mu-1)\Delta]}{[\mu\Delta-\tanh(v)][\Delta+\tanh(v)]}\right\}P_s(v).
\end{eqnarray}
By integrating Eq.~(\ref{apeqn16}), we derive the following expression for $P_s(v)$:
\begin{eqnarray}
\label{apeqn17}
P_s(v) = \frac{N}{(\mu\Delta - \tanh(v))(\Delta + \tanh(v))} 
\exp \left\{-\lambda\int_{}^{v} dv'\frac{2\tanh(v')-(\mu-1)\Delta}{(\mu\Delta - \tanh(v'))(\Delta + \tanh(v'))} \right\}.
\end{eqnarray}
We perform the integration in Eq.~(\ref{apeqn17}), which leads to the following expression for $P_s(v)$:
\begin{eqnarray}
\label{apeqn18}
P_s(v) = \frac{N}{(\mu\Delta - \tanh(v))(\Delta + \tanh(v))}\frac{\left|\mu\Delta-\tanh(v)\right|^{\frac{\lambda}{1-\mu^2\Delta^2}}\left| \Delta+\tanh(v)\right|^{\frac{\lambda}{1-\Delta^2}}}{\left| 1-\tanh(v)\right|^{\frac{\lambda(2-(\mu-1)\Delta)}{2(1-\mu\Delta)(1+\Delta)}}\left| 1+\tanh(v)\right|^{\frac{\lambda(2+(\mu-1)\Delta)}{2(1+\mu\Delta)(1-\Delta)}}}.
\end{eqnarray}
The integration constant $N$ cannot be determined analytically because the integration of $P_s(v)$ does not have a closed-form solution. Additionally, it is evident from the expression of $P_s(v)$ that it is not a symmetric under the transformation $v\rightarrow -v$ when $\mu\ne 1$. For $\mu=1$, Eq.~(\ref{apeqn18}) reduces to
\begin{eqnarray}
\label{apeqn18a}
P_s(v) = \frac{N}{\Delta^2 -\tanh^2( v)} 
\left| \frac{\Delta^2- \tanh^2( v)}{1 - \tanh^2(v)} \right|^{\frac{\lambda}{(1-\Delta^2)}}.
\end{eqnarray}
Clearly, the expression of $P_s(v)$ in Eq.~(\ref{apeqn18a}) is symmetric under the transformation $v\rightarrow -v$. Thus, we conclude that $P_s(v)$ is symmetric only when the active force is also symmetric. These expressions of $P_s(v)$ will assist in validating the numerically computed $P_s(v)$ for various active force parameters.

\begin{figure}
\centering
\includegraphics*[width=0.86\textwidth]{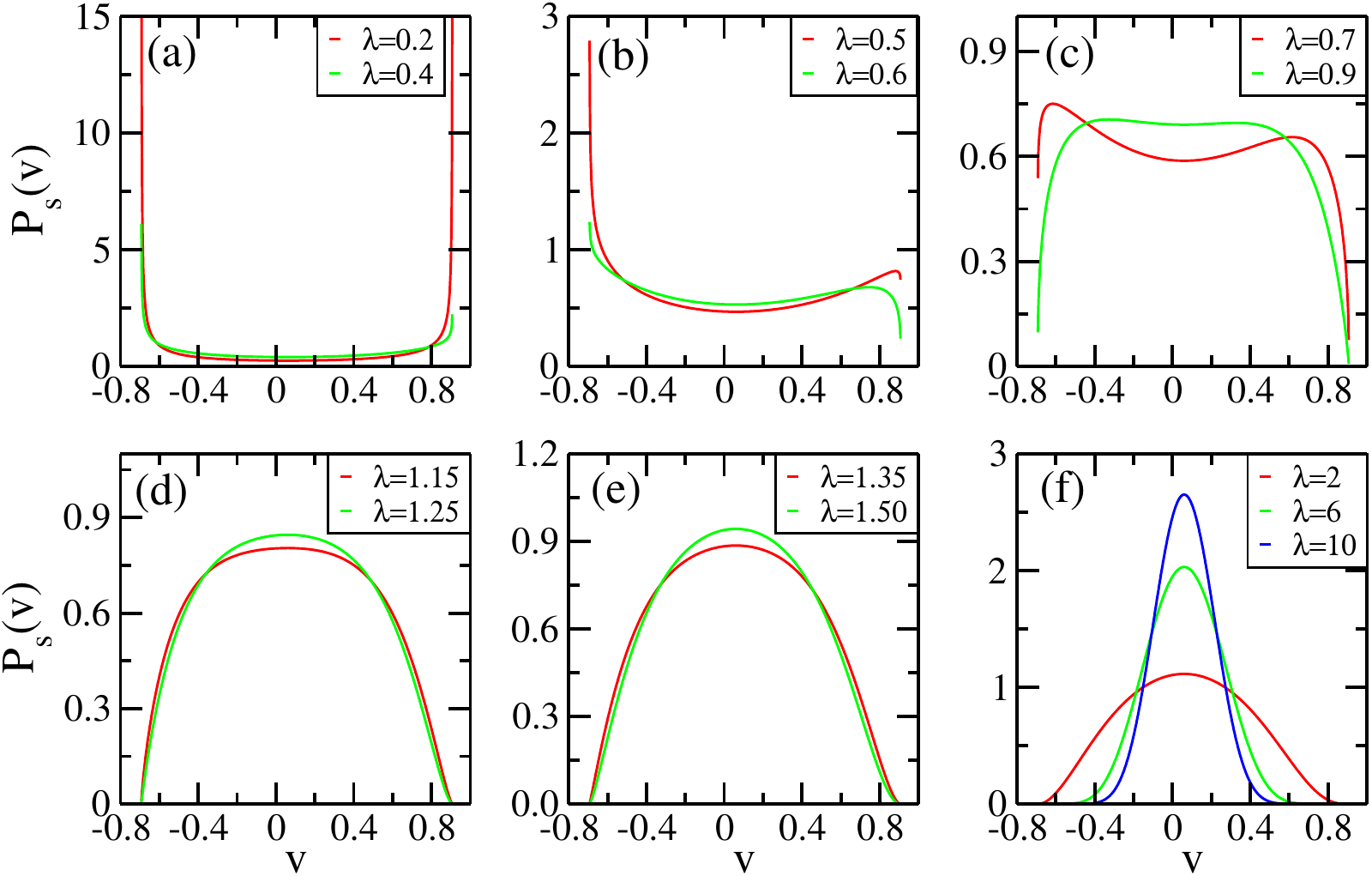}
\caption{\label{fig1a} Plot of the normalized steady-state velocity distribution function $P_s(v)$ for $\Delta=0.6$ and $\mu=1.2$ with the following ranges of $\lambda$: (a) $\lambda\in(0.2,0.4)$, (b) $\lambda\in(0.5,0.6)$, (c) $\lambda\in(0.7,0.9)$, (d) $\lambda\in(1.15,1.25)$, (e) $\lambda\in(1.35,1.50)$, and (f) $\lambda\in(2, 10)$, as indicated.}
\end{figure}

We numerically compute $N$ to normalize the steady-state velocity distribution, $P_s(v)$. First, we consider the asymmetric active force ($\mu\ne 1$) case. Figure~\ref{fig1a} presents the normalized plot of $P_s(v)$ for $\Delta=0.6$ and $\mu=1.2$ at different values of $\lambda$, as specified. In this case, the particle's velocity is constrained within the range $(-\tanh^{-1}{\Delta}, +\tanh^{-1}\mu\Delta)$, making $P_s(v)$ asymmetric about $v=0$. Notably, several intriguing transitions in $P_s(v)$ are observed as $\lambda$ is varied due to the dynamical trapping of particle velocity. In particular, at $v_a$, $P_s(v)$ diverges to infinity for $\lambda < 0.48$ (approx.), whereas for $\lambda > 0.48$, it approaches zero, while the divergence at $v_b$ remains unchanged, as shown in Figs.~\ref{fig1a}(a)-(b). When $\lambda < 0.48$, the particle can attain the maximum possible velocities $v_a$ and $v_b$, leading to the divergence of both $P_s(v_a)$ and $P_s(v_b)$. However, for $\lambda > 0.48$, the particle velocity can still reach $v_b$ but not $v_a$ due to $\mu > 1$, resulting in $P_s(v_a) \to 0$ and $P_s(v_b) \to \infty$. For $0.7 < \lambda < 0.9$, $P_s(v)$ approaches zero at both $v_a$ and $v_b$ vertically, while two peaks emerge within the interval $v\in[v_b,v_a]$, as shown in Fig.~\ref{fig1a}(c). In this range of $\lambda$, particle velocity stays close to $v_a$ or $v_b$, leading to a rapid suppression of $P_s(v_a)$ and $P_s(v_b)$ to zero, while peaks in $P_s(v)$ emerge near $v_a$ and $v_b$. Additionally, for $1.15 < \lambda < 1.25$, $P_s(v)$ approaches zero vertically at $v_b$ and horizontally at $v_a$, whereas for $1.35 < \lambda < 1.5$, it vanishes horizontally at both $v_b$ and $v_a$, as shown in Figs.~\ref{fig1a}(d) and (e), respectively. $P_s(v)$ approaches zero horizontally at $v_a$, $v_b$, or both when the maximum velocity the particle can attain remains well below $v_a$, $v_b$, or both for a given $\lambda$. For $\lambda\approx 1$, the two maxima of $P_s(v)$ merge, forming a bell-shaped curve. The peak location, $v=v_p$, of the bell-shaped curve depends on $\Delta$ and $\mu$. In this case, the particle velocity mainly fluctuates around $v_p$. Figure~\ref{fig1a}(f) shows $P_s(v)$ for sufficiently large values of $\lambda$, showing a decrease in its width as $\lambda$ increases.

\begin{figure}
\centering
\includegraphics*[width=0.85\textwidth]{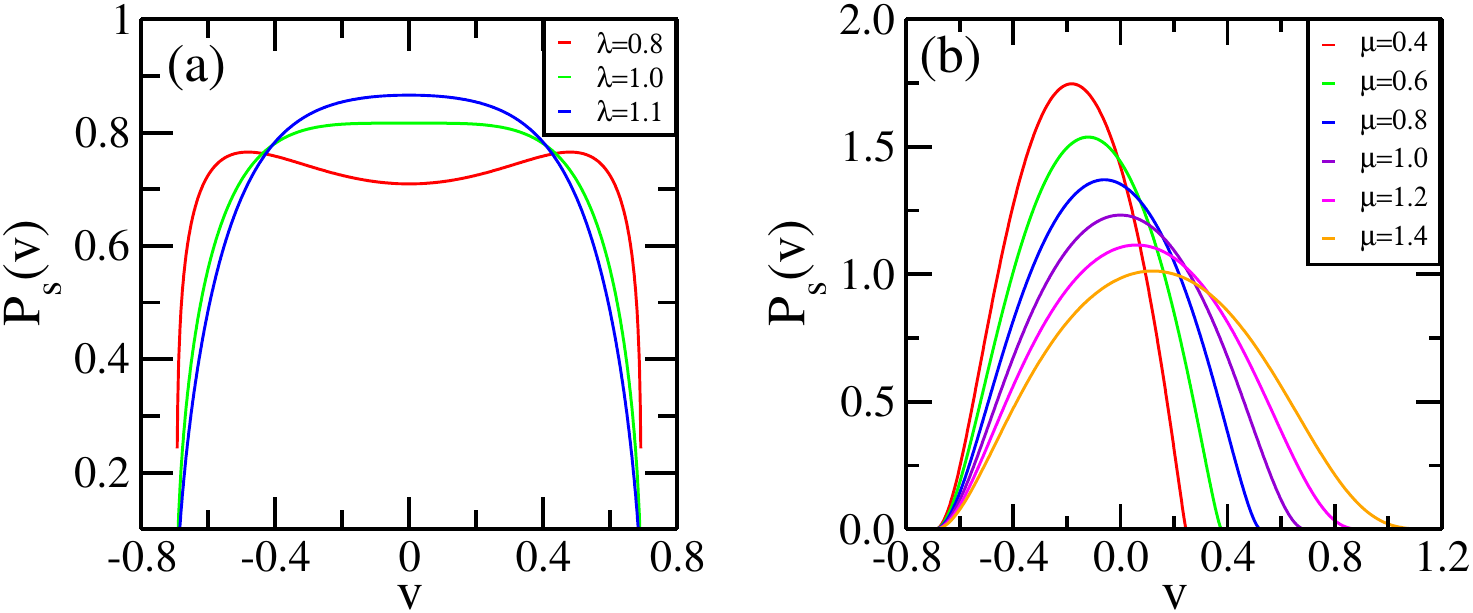}
\caption{\label{fig1b} Normalized steady-state velocity distribution \( P_s(v) \) for (a)  $\Delta=0.6$, $\mu=1$, and varying $\lambda$, and (b)  $\Delta=0.6$, $\lambda=1$, and different values $\mu$, as indicated.}
\end{figure}

When $\mu=1$, i.e., in the symmetric active force case, $P_s(v)$ is symmetric about $v=0$, as given by eq.~(\ref{apeqn18a}). Consequently, for $\lambda<1$, we observe that $P_s(v)$ exhibits two maxima of equal height, as shown in Fig.~\ref{fig1b}(a) for $\lambda=0.8$. For $\lambda\ge 1$, these two maxima merge at $v=0$, resulting in a bell-shaped curve with a single peak at  $v=0$. Furthermore, for $\mu=1$ and a fixed $\lambda$, $P_s(v)$ shows identical behavior at $v=v_a$ and $v=v_b$.

In Fig.~\ref{fig1b}(b), we plot the normalized $P_s(v)$ for $\Delta=0.6$, $\lambda=2$, and various values of $\mu$, as mentioned. In this case $v_a$ can be either smaller or larger than $|v_b|$, depending on $\mu$. It is clear that the peak of $P_s(v)$ lies in $v>0$ when $v_a>|v_b|$ and in $v<0$ when $v_a <|v_b|$. For $\mu=1$, the peak of $P_s(v)$ remains at $v=0$.

\subsection{Activity-Induced Transition in $P_s(v)$}\label{2b}
We examine the behavior of $P_s(v)$ at $v=v_a$ and $v=v_b$, the fixed points or attractors of velocity dynamics of the active particle. First, we focus on the behavior at $v=v_b$. The Taylor expansion of $P_s(v)$ around $v=v_b$ yields
\begin{eqnarray}
\label{apeqn19}
P_s(v) \sim \left| v - v_b \right|^{\frac{\lambda}{(1-\Delta^2)}-1}.
\end{eqnarray}
Clearly, it depends on two parameters: the flipping rate $\lambda$ between the two active force states and the strength of the active force $\Delta$.

In Eq.~(\ref{apeqn19}), $P_s(v)$ diverges to infinity at $v=v_b$ in an integrable manner when
\begin{equation}
\label{apeqn20}
\frac{\lambda}{1 - \Delta^2} < 1.
\end{equation}
This condition is achievable when $\lambda<1-\Delta^2$. Since $|\Delta|\le 1$, the flipping rate $\lambda$ is always less than unity in this case. As a result, the active noise has sufficient time to drive the particle velocity toward the attractor at $v=v_b$.

When $\lambda$ is large, we arrive at the condition
\begin{equation}
\label{apeqn21}
\frac{\lambda}{1 - \Delta^2} - 1>0.
\end{equation}
In this case, the flipping dynamics of the active force efficiently suppresses the influence of the attractors at $v=v_b$. As a result, $P_s(v)$ approaches zero at $v=v_b$, either with a divergent slope or a zero slope. To examine the precise nature of this behavior, we differentiate Eq.~(\ref{apeqn19}), which yields:
\begin{eqnarray}
\label{apeqn22}
P_s^{\prime}(v)\Big|_{v=v_b} \sim \left| v - v_b \right|^{\frac{\lambda}{(1-\Delta^2)}-2}.
\end{eqnarray}
Clearly, $P_s^{\prime}(v_b)\rightarrow\infty$ if $1-\Delta^2>\lambda/2$. Therefore, for the range $\lambda>1-\Delta^2>\lambda/2$, $P_s(v_b)\rightarrow 0$ while $P_s^{\prime}(v_b)\rightarrow\infty$. This indicates a cusp-like behavior at $v=v_b$, where the probability density $P_s(v)$ sharply decreases to zero but with an infinite slope. Again, $P_s^{\prime}(v_b)\rightarrow 0$ if $1-\Delta^2<\lambda/2$. Therefore, for $1-\Delta^2<\lambda/2$, we observe that both $P_s(v_b)$ and $P_s^{\prime}(v_b)$ approaches to zero.

The behavior of $P_s(v)$ near $v=v_a$ can be determined using the above analysis by substituting $\Delta$ with $\mu\Delta$, leading to the following results: \ \\ (a) $P_s(v_a)\rightarrow\infty$ when $\lambda< 1- (\mu\Delta)^2$;\ \\ (b) For $\lambda>1-(\mu\Delta)^2>\lambda/2$, $P_s(v_a)\rightarrow 0$ while $P_s^{\prime}(v_a)\rightarrow\infty$; \ \\ (c) When $1-(\mu\Delta)^2<\lambda/2$, both $P_s(v_a)$ and $P_s^{\prime}(v_a)$ approach zero.\ \\ These findings are consistent with the results shown in Fig.~\ref{fig1a} for $\Delta=0.6$ and $\mu=1.2$, where the transition points are calculated as follows: \ \\ i) $\lambda_1=1-\mu^2\Delta^2=0.4816$ (Fig.~\ref{fig1a}(b)), above which $P_s(v_a)\rightarrow 0$ and $P_s^\prime(v_a)\rightarrow \infty$ while $P_s(v_b)\rightarrow \infty$.\ \\
ii) $\lambda_2=1-\Delta^2=0.64$ (Fig.~\ref{fig1a}(c)), above which both $P_s(v_a), P_s(v_b)\rightarrow 0$,  and $P_s^\prime(v_a), P_s^\prime(v_b)\rightarrow \infty$. \ \\ 
iii) $\lambda_3=2(1-\mu^2\Delta^2)=0.9632$ (Fig.~\ref{fig1a}(d)), above which both $P_s(v_a), P_s(v_b)\rightarrow 0$, but $P_s^\prime(v_a)\rightarrow 0$ while $P_s^\prime(v_a)\rightarrow \infty$.\ \\ 
iv) $\lambda_4=2(1-\Delta^2)=1.28$ (Fig.~\ref{fig1a}(e)), above which  $P_s(v_a)$, $P_s(v_b)$, $P_s^\prime(v_a)$ and $P_s^\prime(v_b)$ all approach zero.

Next, we identify the extrema of $P_s(v)$. To determine the velocities $v = u$ at which these extrema occur, we differentiate Eq.~(\ref{apeqn18}) with respect to $v$, yielding the following relation:
\begin{eqnarray}
\label{apeqn23}
\left(1-\tanh^2(u)\right)\left\{\frac{\tanh(u)}{\lambda}+\frac{(\mu-1)\Delta}{4\lambda}\right\}-\tanh(u)+\frac{(\mu-1)\Delta}{2}=0
\end{eqnarray}
For $\mu=1$, the roots of eq.~(\ref{apeqn23}) are at
\begin{eqnarray}
\label{apeqn24}
u_0 = 0 \quad \text{and}\quad u_{1,2} = \pm  \tanh^{-1}\left(\sqrt{1 - \lambda}\right).
\end{eqnarray}
This reveals the following scenarios:\ \\
(a) $\lambda < 1 - \Delta^2$: In this regime, the root at $v = u_0=0$ is the minimum, and $v = \pm v_b$ and are the maxima of $P_s(v)$.\ \\
(b) $1 - \Delta^2 < \lambda < 1$: Here, the root at $v = u_0=0$ remains the minimum, but additional $\lambda$-dependent roots $v = u_{1,2}$ emerge as the maxima of $P_s(v)$.\ \\
(c) $\lambda > 1$: $v = u_0=0$ becomes the only maximum of $P_s(v)$.\ \\
These findings align with the results shown in Fig.~\ref{fig1b}(a) for $\Delta=0.6$.

For $\mu\ne 1$, $P_s(v)$ can exhibit two maxima and one minimum within the interval $v\in[v_b,v_a]$ when $\max(1-\Delta^2,1-\mu^2\Delta^2)<\lambda\lesssim 1$. In this range of $\lambda$, both $P_s(v_a)$ and $P_s(v_b)$ approach zero. The exact locations of these extrema can be determined by solving eq.~(\ref{apeqn23}). For $\lambda>1$, the two maxima merge into a single peak. In the limit $\lambda\rightarrow\infty$, the position of this peak is given by  
\begin{eqnarray}
\label{apeqn24a}
u_a=\tanh^{-1}{\left[\frac{(\mu-1)\Delta}{2}\right]}
\end{eqnarray}
Clearly, for $\mu<1$, $u_a<0$, meaning the peak of $P_s(v)$ lies in the negative $v$ domain, whereas for $\mu>1$, $u_a>0$, indicating that the peak lies in the positive $v$ domain.

\subsection{Effective Diffusion Coefficient $D_e$ for $\mu=1$}\label{sec2c}
From Eq.~(\ref{apeqn24a}), it is clear that $\mu\neq 1$ introduces a nonzero drift in the particle's velocity. Consequently, the particle exhibits purely diffusive motion only when $\mu=1$. Here, we calculate the effective diffusion coefficient $D_e$ of particle motion to gain a deeper understanding of the transport properties of the active particle in a shear-thinning environment for $\mu=1$. Using the Kubo relation, $D_e$ can be expressed as an integral over velocity auto correlation function~(VACF) as follows:
\begin{eqnarray}
\label{apeqn25}
D_{e} = \int_0^\infty dt \, \left[\langle v(0)v(t) \rangle - \langle v(0) \rangle^2\right]
\end{eqnarray}
in $d=1$. Since, the velocity distribution function for $\mu=1$ is an even function of $v$ (eq.~(\ref{apeqn18a})), $\langle v(0) \rangle=0$. Therefore, eq.~(\ref{apeqn25}) reduces to
\begin{eqnarray}
\label{apeqn26}
D_{e} = \int_0^\infty dt \, \langle v(0)v(t) \rangle.
\end{eqnarray}
The exact quadrature expression for the diffusion coefficient of a particle driven by a symmetric dichotomous noise and subject to a general odd friction function was previously derived by Lindner~\cite{B10}. Using that method, we derive the following expression for $D_e$:
\begin{eqnarray}
\label{apeqn27}
D_{\text{e}} = 2\lambda \Delta^2 \frac{\int_0^{v_a} dx \, e^{V(x)} \left[ \Delta^2 - \tanh^2(x) \right]^{-1}       \left\{\int_x^{v_a} dy \,y e^{-V(y)} \left[ \Delta^2 -  \tanh^2(y) \right]^{-1}\right\}^2}{\int_0^{v_a} dz \, e^{-V(z)} \left[\Delta^2 - \tanh^2(z) \right]^{-1}},
\end{eqnarray}
where the function $V(x)$ has the following form:
\begin{eqnarray}
\label{apeqn28}
V(x) = 2\lambda \int_0^v dx \, \frac{\tanh(x)}{\Delta^2 - \tanh^2(x)}.
\end{eqnarray}
Clearly, $D_e$ depends on the active force parameters $\Delta$ and $\lambda$. Since the exact nature of this dependence cannot be determined analytically due to the absence of closed-form solutions for the integrals, we evaluate it computationally in Sec.~\ref{sec3}.

\section{Numerical Results}\label{sec3}
We numerically solve Eq.~(\ref{apeqn1b}) using the Euler-discretization scheme to update the particle's velocity and compute the $P_s(v)$, mean-square velocity $\langle v^2\rangle(t)$, and velocity auto-correlation function $\langle v(0)v(t)\rangle$ in $d=1$. The discretized version of Eq.~(\ref{apeqn1b}) is given by:
\begin{eqnarray}
\label{apeqn31}
v(t+dt)=v(t) - [\tanh v(t) - \xi(t)] dt.
\end{eqnarray}
The position of the particle at time $t$, denoted as $r(t)$, is updated using the following discretized equation:
\begin{eqnarray}
\label{apeqn32}
r(t+dt)=r(t) + v(t) dt.
\end{eqnarray}
Here, $dt$ is the discretized time step. We generate the dichotomous noise $\xi(t)$ following the method outlined in Ref.~\cite{CE06}. The probability $p$ that $\xi(t)$, starting from state $\mu\Delta$(or $-\Delta$), transitions to state $-\Delta$(or $\mu\Delta$) within a time interval $dt$ is given by 
\begin{eqnarray}
\label{apeqn33}
p=\frac{1}{2}\{1-\exp(-2\lambda dt)\}.
\end{eqnarray}
Next, we generate a random number $w$ in the interval [0,1] using a uniform random number generator. When $p>w$, $\xi(t)$ transitions from one state to another. In the simulation, the numerical value of $dt=0.001$ is used and the results are obtained by averaging over $10^5$ trajectories. At $t=0$, the particle is initialized with zero velocity and positioned at the origin. The velocity and position are updated iteratively using Eqs.~(\ref{apeqn31}) and (\ref{apeqn32}), respectively.

\begin{figure}
\centering
\includegraphics*[width=0.78\textwidth]{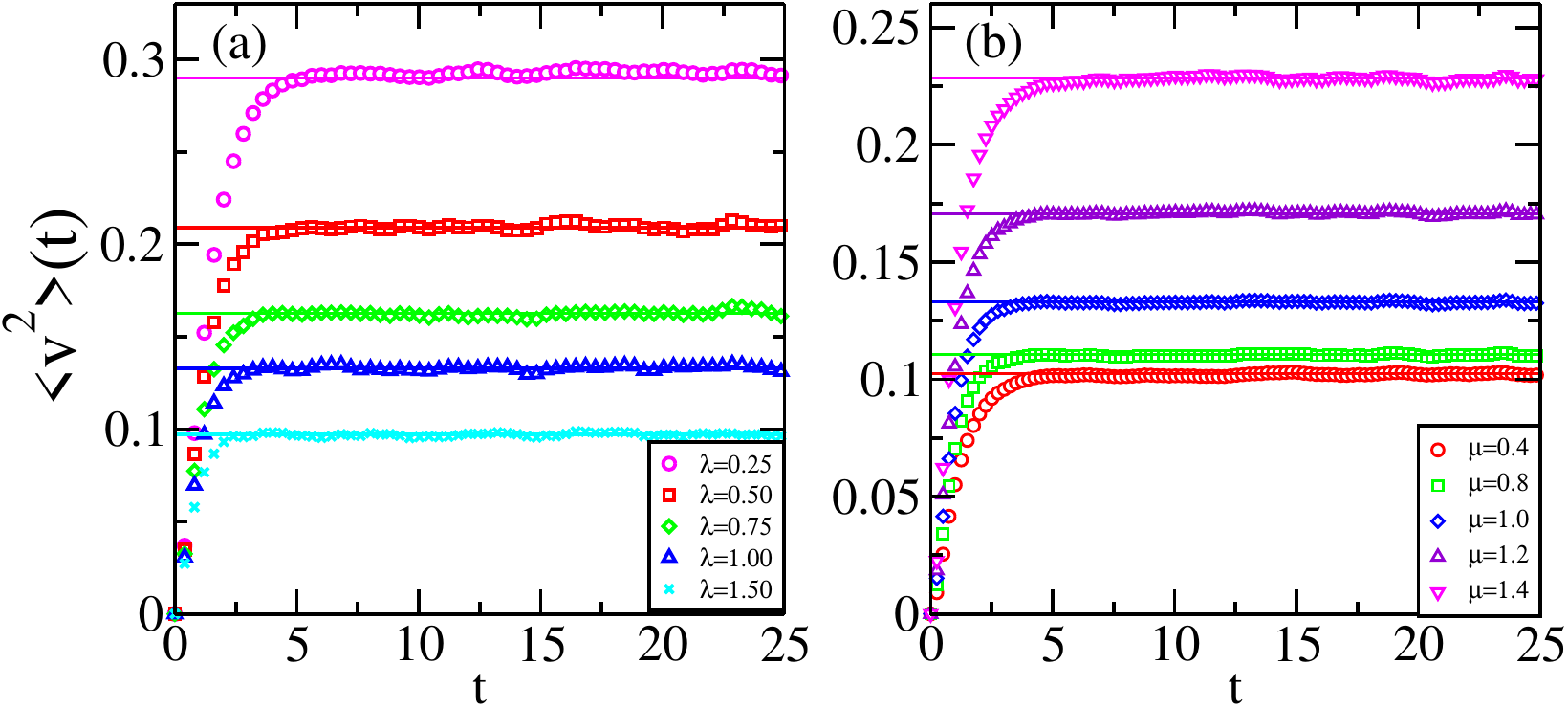}
\caption{\label{fig2} Plot of the mean-squared velocity, $\langle v^2\rangle(t)$, as a function of time $t$ for $\Delta=0.6$: (a) $\mu=1$ with different values of $\lambda$, and (b) $\lambda=1$ with different values of $\mu$, as specified. The data points represent numerical results, while the solid lines correspond to the steady-state values of mean-squared velocity $\langle v^2\rangle_s$ derived from Eq.~(\ref{apeqn34}). The data points and the solid line of a given color are obtained for the same values of $\Delta$, $\mu$, and $\lambda$.}
\end{figure}

Figure~\ref{fig2}(a) shows the plot of $\langle v^2\rangle(t)$ vs. $t$ for $\Delta=0.6$, $\mu=1$, and different values of $\lambda$, as mentioned. After an early transient, $\langle v^2\rangle(t)$ reaches a steady-state value for all cases. We then numerically evaluate steady-state mean-squared velocity $\langle v^2\rangle_s$ by solving the following equation:
\begin{eqnarray}
\label{apeqn34}
\langle v^2\rangle_s = \frac{\int_{v_b}^{v_a} v^2 P_s(v) dv}{\int_{v_b}^{v_a}P_s(v) dv},
\end{eqnarray}
where $P_s(v)$ is given by eq.~(\ref{apeqn18}). Clearly, the long-time numerical values of $\langle v^2\rangle(t)$ agree with the numerically computed $\langle v^2\rangle_s$ obtained from eq.~(\ref{apeqn34}) for all cases. Additionally, for a fixed $\mu$ and $\Delta$, the time $t_s$ at which the particle reaches $\langle v^2\rangle_s$ depends on $\lambda$. A higher $\lambda$ results in a smaller $t_s$. This occurs because more frequent flipping of the active force accelerates the approach to the steady state.

Figure~\ref{fig2}(b) shows the plot of $\langle v^2\rangle(t)$ vs. $t$ for $\Delta=0.6$, $\lambda=1$, and various values of $\mu$, as specified. Similar to the previous case, the numerical values of $\langle v^2\rangle(t)$ at long times show excellent agreement with the numerically computed $\langle v^2\rangle_s$ obtained from eq.~(\ref{apeqn34}) for all cases.

\begin{figure}
\centering
\includegraphics*[width=0.80\textwidth]{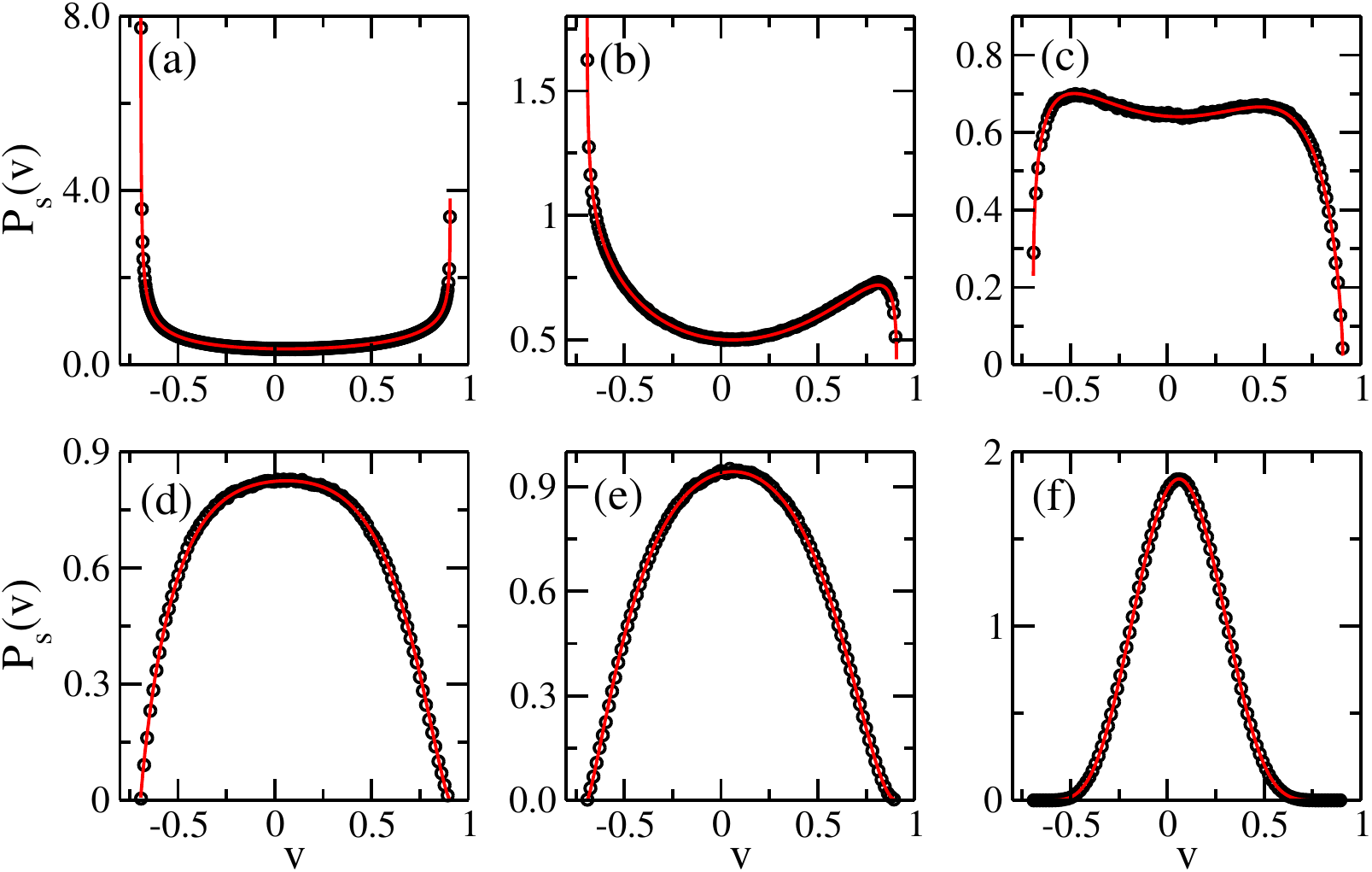}
\caption{\label{fig3} Plot of the normalized steady-state velocity distribution function $P_s(v)$ for $\Delta=0.6$, $\mu=1.2$, and different values of $\lambda$: (a) $\lambda=0.35$, (b) $\lambda=0.55$, (c) $\lambda=0.8$, (d) $\lambda=1.2$, (e) $\lambda=1.5$, and (f) $\lambda=5$. The black circles represent numerical data, while the solid red lines correspond to the normalized form of $P_s(v)$ as given in Eq.~(\ref{apeqn18}).}
\end{figure}

Next, we compute the normalized steady-state velocity distribution $P_s(v)$ by collecting the velocities of particles and constructing a histogram from the collected data. Velocities were collected during the time interval $t\in (300,400)$ for all the $10^5$ trajectories. Since $dt=0.001$, each histogram was constructed using $10^{10}$ data points. Figure~\ref{fig3} shows the plot of the numerically computed and analytical (normalized) $P_s(v)$s for $\Delta=0.6$, $\mu=1.2$, and various values of $\lambda$, as indicated. It is evident that the numerical and analytical $P_s(v)$ are indistinguishable. Thus, we conclude that the numerical data agrees well with the analytical results. Also, this result holds for other values of $\Delta$, $\mu$, and $\lambda$.

\begin{figure}
\centering
\includegraphics*[width=0.9\textwidth]{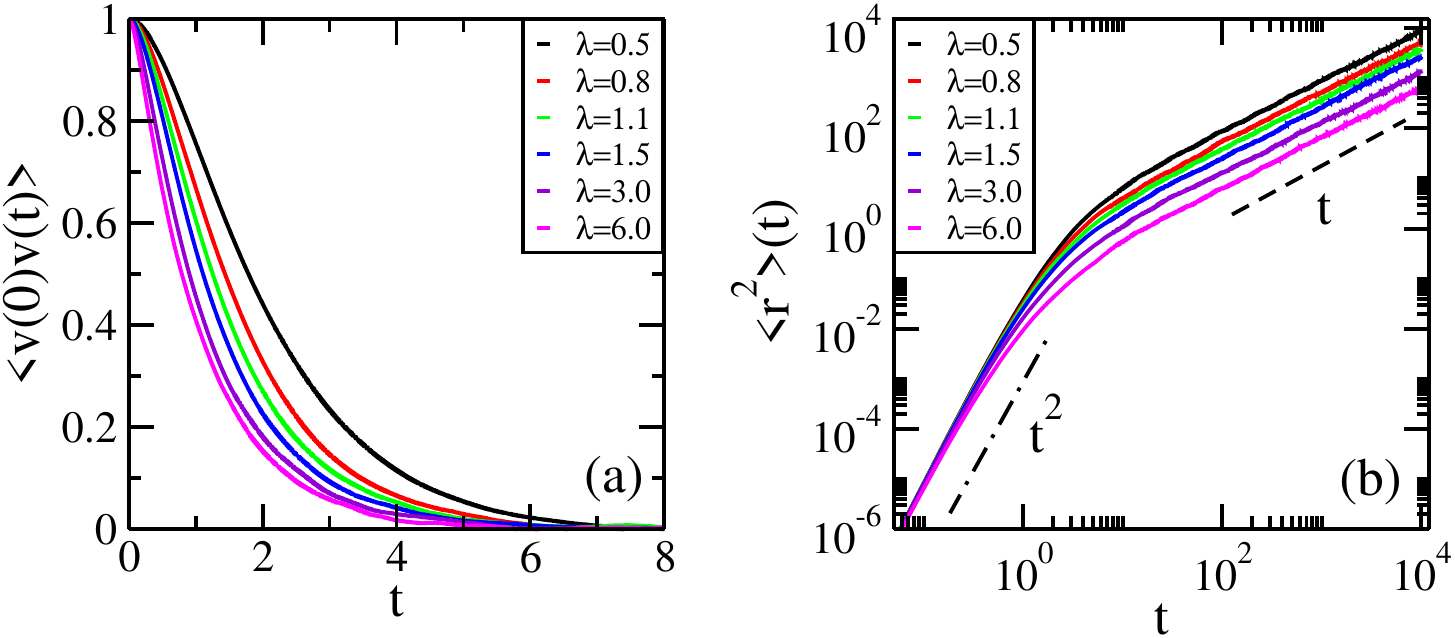}
\caption{\label{fig4} (a) Plot of the velocity autocorrelation function $\langle v(0)v(t)\rangle$ vs. $t$  for $\Delta=0.6$, $\mu=1$, and various values of $\lambda$, as indicated. (b) Plot of the mean-squared displacement $\langle r^2\rangle(t)$ vs. $t$ on a log-log scale for the same values of $\Delta$ and $\lambda$ as used in(a). The dashed-dotted line labeled $t^2$ and the solid-dashed line labeled $t$ represent the ballistic and diffusive regimes, respectively.}
\end{figure}

We numerically compute the velocity auto-correlation function, $C(t)=\langle v(0)v(t)\rangle$ in the steady-state and the mean-squared displacement $\langle r^2\rangle(t)$ for $\mu=1$ to gain insights into the transport properties of the system. Figure~\ref{fig4}(a) shows the plot of $C(t)$ vs. $t$ for $\Delta=0.6$ and different values of $\lambda$, as mentioned. It is apparent that $C(t)$ decays to zero more rapidly with increasing $\lambda$, indicating that the area under the $C(t)$ curve decreases as $\lambda$ increases. Consequently, we conclude that the effective diffusion coefficient $D_e$ decreases as $\lambda$ increases. This conclusion is further supported by examining the variation of $\langle r^2\rangle(t)$ as a function of $t$. In Fig.~\ref{fig4}(b), we plot $\langle r^2\rangle(t)$ vs. $t$ on a log-log scale for $\Delta=0.6$, $\mu=1$, and different values of $\lambda$, as mentioned. At early times, $\langle r^2\rangle(t)\sim t^2$, corresponding to the ballistic regime. This superdiffusive behavior is a natural consequence of the initial deterministic motion of the particle before friction dominates and the velocity of the particle becomes fully randomized by the active force. However, at later times, the particles exhibit diffusive motion, where $\langle r^2\rangle(t)\sim t$. The decreasing slopes of the curves in the diffusive regime with increasing $\lambda$ confirm that $D_e$ decreases with increasing $\lambda$ for a fixed $\Delta$ and $\mu=1$. Physically, at higher $\lambda$, the frequent flipping of the active force prevents the particle from reaching its maximum possible velocities. As a result, the particle's velocity primarily fluctuates around $v=0$, reducing the mean-squared displacement and consequently decreasing $D_e$.

Finally, we focus on the variation of $D_e$ as a function of activity parameters $\Delta$ and $\lambda$ for the symmetric case, i.e., $\mu=1$. First, we numerically compute $D_e$ by integrating $C(t)$ over time (see Eq.~(\ref{apeqn26})) and from the slope of the $\langle r^2\rangle(t)$ vs. $t$ curves shown in Fig.~\ref{fig4}(b). For a given $\Delta$ and $\lambda$, the values of $D_e$ obtained using both methods are identical within numerical accuracy. Then we compute $D_e$ by evaluating the integrals in Eq.(\ref{apeqn27}) for various values of $\Delta$ and $\lambda$~\cite{B10}. A detailed numerical approach for solving the quadrature expression in Eq.~(\ref{apeqn27}) is presented in the appendix of Ref.~\cite{B10}. Figure~\ref{fig5} shows the plots of (a) $D_e$ vs. $\lambda$ for various values of $\Delta$ and (b) $D_e$ vs. $\Delta$ for different values of $\lambda$ on a log-log scale. In both cases, $D_e$ obtained from the former method is in good agreement with the value obtained from the latter method. For $\mu=1$ and a fixed $\lambda$, $D_e$ increases with $\Delta$ since a larger $\Delta$ expands the range of achievable particle velocities. This increases the mean-squared displacement and consequently increasing $D_e$.
\begin{figure}
\centering
\includegraphics*[width=0.96\textwidth]{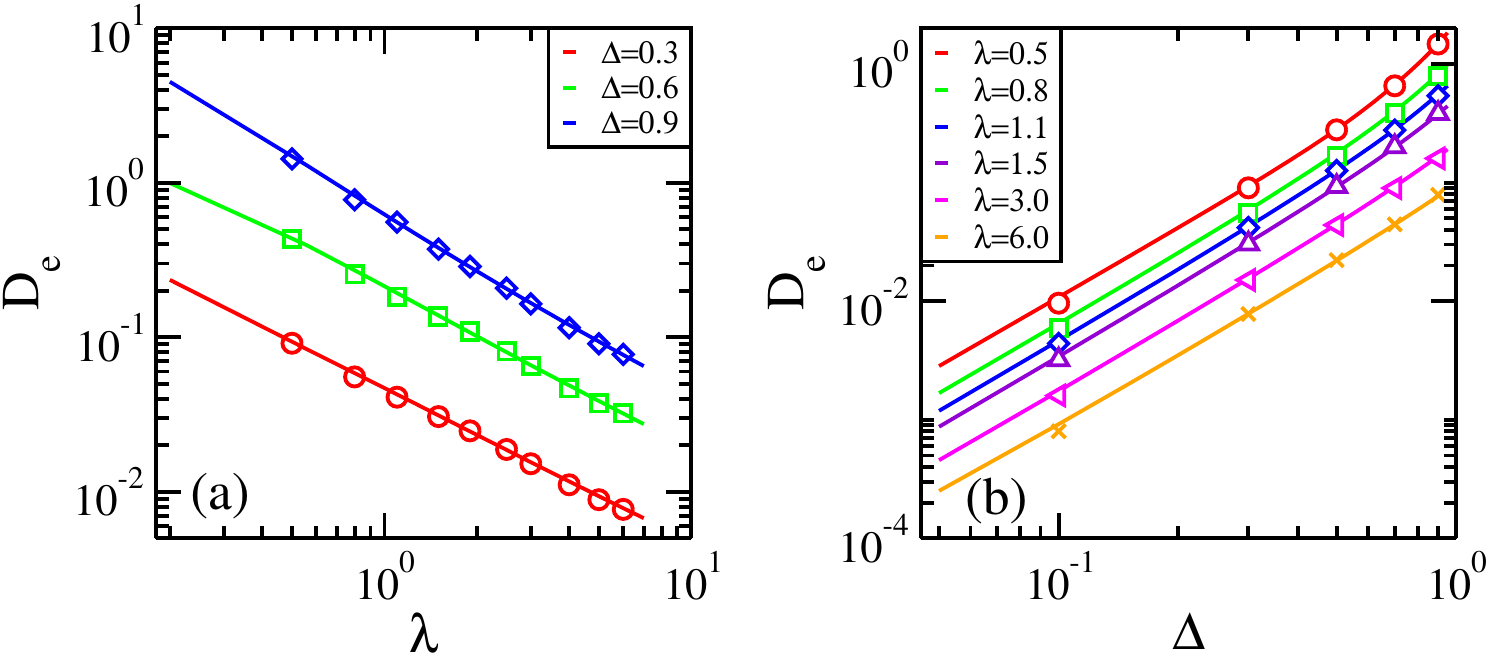}
\caption{\label{fig5} Plot of the effective diffusion coefficient $D_e$ for $\mu=1$ at various $\Delta$ and $\lambda$. Log-log plot of (a) $D_e$ vs. $\lambda$ for various values of $\Delta$ and (b) $D_e$ vs. $\Delta$ for different values of $\lambda$, as mentioned. The numerical data obtained by integrating the velocity autocorrelation function are represented by points, while the lines correspond to $D_e$ computed numerically using Eq.~(\ref{apeqn27}). Lines and points of the same color represent $D_e$ for the same values of $\Delta$ and $\lambda$.}
\end{figure}

\section{Universal Nature of Transitions in $P_s(v)$}\label{sec4}
\begin{figure}
\centering
\includegraphics*[width=0.86\textwidth]{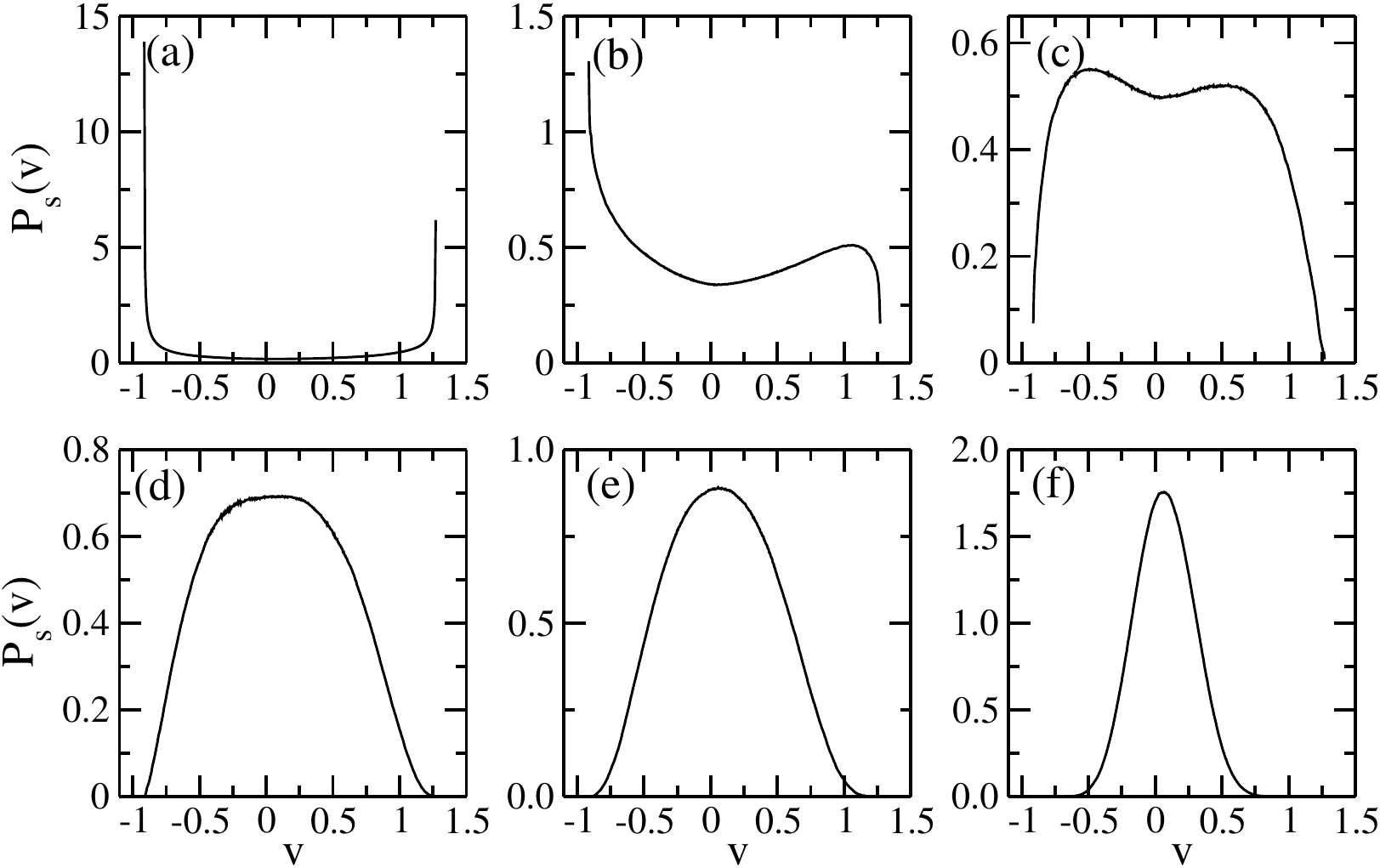}
\caption{\label{fig6} Numerically obtained plots of the normalized steady-state velocity distribution function $P_s(v)$ for $F_f(v)$ given by Eq.~(\ref{apeqn41}) at $\Delta = 0.6$, $\mu=1.2$ and various values of $\lambda$: (a) $\lambda = 0.15$, (b) $\lambda = 0.35$, (c) $\lambda = 0.6$, (d) $\lambda = 1$, (e) $\lambda = 1.5$, and (f) $\lambda = 5$, as indicated.} 
\end{figure}

In Section~\ref{2b}, we observed that $P_s(v)$ exhibits multiple transitions in the vicinity of $v_a$ and $v_b$, influenced by the active force parameters $\Delta$ and $\lambda$, when the viscous force from the medium is modeled as $F_f(v) \sim \tanh(v)$. This observation raises an important question: Are these transitions universal across different choices of $F_f(v)$ representing the shear-thinning medium? To explore this, we consider an alternative model based on the modified Tustin empirical form~\cite{ALACH24}, where the viscous force in a shear-thinning medium is given by
\begin{eqnarray}
\label{apeqn41}
F_f(v) \sim \left(1-\exp\left(-\frac{|v|}{v_0}\right)\right)\frac{v}{|v|}.
\end{eqnarray}
This expression ensures that $F_f(v)$ saturates to a constant at large $v$, thereby effectively capturing the shear-thinning nature of the medium. In this case, the Langevin equation for the particle's motion in $d=1$ becomes
\begin{eqnarray}
\label{apeqn42}
m\frac{dv}{dt} = -\gamma\left(1-\exp\left(-\frac{|v|}{v_0}\right)\right)\frac{v}{|v|} + \zeta(t).
\end{eqnarray}
Here, all variables have the usual meanings as described in Sec.~\ref{sec2}. We non-dimensionalize  Eq.~(\ref{apeqn42}) by using the same rescaling of variables given in Sec.~\ref{sec2}, which yields
\begin{eqnarray}
\label{apeqn43}
\frac{dv}{dt} = -\left(1-e^{-|v|}\right)\frac{v}{|v|} + \xi(t),
\end{eqnarray}
In this case, the particle’s velocity always lies within the range $v_b = \ln{(1-\Delta)}$ to $v_a = -\ln{(1-\mu\Delta)}$. We numerically solve Eq.~(\ref{apeqn43}) using the Euler discretization method with a time step of $dt = 0.001$ to compute the time-dependent velocity of the particle, $v(t)$. The system is equilibrated up to $t = 100$, after which $v(t)$ is collected over the interval $t \in (100, 200)$ for $10^5$ independent trajectories to compute $P_s(v)$. Figure~\ref{fig6} shows the plots of $P_s(v)$ for $\Delta=0.6$, $\mu=1.2$ and six different values of $\lambda$, as mentioned. Notably, $P_s(v)$ exhibits all the characteristic behaviors near $v_a$ and $v_b$ that were previously observed in Fig.~\ref{fig1a}. This confirms that the qualitative features of $P_s(v)$ near the extremal velocities are universal and do not depend on the specific form of the viscous force, thereby highlighting the universal nature of these transitions. However, the precise relationship between $\Delta$ and $\lambda$ that governs the transition points may still vary with the form of the viscous force.

\section{Summary and Discussion}\label{sec5}
Let us conclude this paper with a summary and discussion of our results. We have studied the dynamics of an athermal inertial run-and-tumble particle moving in a shear-thinning medium, where a Coulomb-tanh function models the viscosity of the medium. The activity is represented by an asymmetric dichotomous noise with strengths $-\Delta$ and $\mu\Delta$, transitioning between these states at a rate $\lambda$, which mimics the behavior of a run-and-tumble particle in $d=1$. There are numerous physical and biological systems where motion is effectively one-dimensional. In physics, this includes the transport of particles through nanopores, optical traps, and microfluidic channels where confinement restricts movement along a line. In biological systems, examples range from molecular transport along cytoskeletal filaments to particle diffusion in narrow, confined geometries. Our one dimensional model directly reflects these scenarios, making it not only sufficiently realistic but also physically well-motivated.

We begin with the Langevin equation of motion for the particle and derive its non-dimensional form by rescaling velocity and time using the appropriate scales. Next, using the Fokker-Planck~(FP) equation for the time-dependent probability distribution functions $P(v,-\Delta,t)$ and $P(v,\mu\Delta,t)$ of the particle's velocity $v$ at time $t$, moving under an active force $-\Delta$ and $\mu\Delta$ respectively, we analytically derive the steady-state velocity distribution function $P_s(v)$. We found that the shape of $P_s(v)$ depends on $\Delta$, $\mu$, and $\lambda$. Additionally, $P_s(v)$ is confined within the lower limit $v_b=-\tanh^{-1}{\Delta}$ and the upper limit $v_a=\tanh^{-1}{\mu\Delta}$ of the particle velocity. For $\mu\ne 1$, $P_s(v)$ is an asymmetric function of $v$. In this case, for a given $\Delta$ and $\mu$, $P_s(v)$ undergoes multiple transitions when $\lambda$ is varied: (a) When $\lambda < 1 - \mu^2\Delta^2$, $P_s(v_a)$ diverges; (b) When $\lambda>1-\mu^2\Delta^2>\lambda/2$, $P_s(v_a)\rightarrow 0$ while $P_s^\prime (v_a)$ diverges; (c) For $1-\mu^2\Delta^2<\lambda/2$, both $P_s(v_a)$ and $P_s^\prime (v_a)$ approach zero. A similar behavior in $P_s(v)$ is observed at $v=v_b$, with the transition points obtained by setting $\mu=1$. Furthermore, when $\max(1-\Delta^2,1-\mu^2\Delta^2)<\lambda\lesssim 1$, $P_s(v)$ exhibits two maxima and a minimum within the interval $v\in[v_b,v_a]$. In this range of $\lambda$, both $P_s(v_a)$ and $P_s(v_b)$ approach zero. The exact location of these extrema depends on active fore parameters. For $\lambda>1$, the two maxima merge into a single peak. In the limit $\lambda\rightarrow\infty$, the position of this peak is determined by $\mu$ and $\Delta$. When $\mu=1$, $P_s(v)$ is symmetric about $v=0$. In this case, the behavior of $P_s(v)$ at both velocity limits ($v=\pm \tanh^{-1}{\Delta}$) will be identical. The single peak of $P_s(v)$ observed for $\lambda>1$ will be centered at $v=0$. Here, we want to emphasize that the transitions observed in $P_s(v)$ near $v_a$ and $v_b$ are governed by the active force parameters only. Any function modeling the shear-thinning behavior of the medium will yield qualitatively identical transitions near $v_a$ and $v_b$. The $\lambda$-$\Delta$ relationship that determines the transition points may depend on the specific form of the function modeling the shear-thinning behavior of the medium. Additionally, we obtain an exact quadrature expression for the effective diffusion coefficient $D_e$ for the symmetric case ($\mu=1$), where the particle exhibits pure diffusion. However, determining the exact analytical form of dependence of $D_e$ on $\Delta$ and $\lambda$ remains challenging due to the nonlinear nature of the viscosity.

We numerically solve the Langevin equation of motion of the particle using the Eular discretization method to obtain its time-dependent velocity, $v(t)$ and position, $r(t)$. We then compute the mean-squared velocity, $\langle v^2\rangle(t)$, as a function of time. At later times, $\langle v^2\rangle(t)$ reaches a steady-state value, $\langle v^2\rangle_s$, which agrees with the numerically computed steady-state value using $P_s(v)$. Next, we compute the normalized steady-state velocity distribution function for $\Delta=0.6$, $\mu=1.2$, and various values of $\lambda$, which also matches with the analytical $P_s(v)$ for the corresponding values of $\Delta$, $\mu$, and $\lambda$. These results remain valid for the other permissible values of $\Delta$, $\mu$, and $\lambda$. To compute $D_e$, we evaluate the velocity autocorrelation function, $C(t)=\langle v(0)v(t)\rangle$, and the mean-squared displacement, $\langle r^2\rangle(t)$, for $\mu=1$ and different values of $\Delta$ and $\lambda$. At large $t$, $\langle r^2\rangle(t)\sim t$, indicating diffusive behavior of the particles at later times. We found that $D_e$ decreases as $\lambda$ increases for a fixed $\Delta$, while it increases as $\Delta$ increases for a fixed $\lambda$. Furthermore, for fixed $\Delta$ and $\lambda$, the value of $D_e$ obtained by integrating $C(t)$ over time matches both the value determined from the slope of the $\langle r^2\rangle(t)$ vs. $t$ curve in the diffusive regime and the analytical prediction.

We believe that our study provides an in-depth understanding of the steady-state velocity distribution and transport properties of athermal inertial active run-and-tumble particles moving through a shear-thinning medium in $d=1$. Extending this analysis to higher dimensions and exploring the impact of spatial dimensionality on various dynamical quantities would be an interesting research avenue. A compelling direction for future work would be to examine the time-dependent and steady-state transport properties of various active systems in a shear-thinning medium, including an inertial active Ornstein-Uhlenbeck particle, a run-and-tumble particle in a harmonic trap, a particle with activity modeled by Lévy noise~\cite{zw19}, or a driven active particle. Additionally, recent advances in robotics have led to the development of self-propelled micro- and nanorobots designed to navigate through shear-thinning media, such as biological fluids like mucus and blood, for efficient drug delivery. Given that these robots possess finite mass, our results for $P_s(v)$ and $D_e$ can be directly applied to describe their motion. This is just one of many possible examples from applied sciences where our study is relevant. We hope that our findings will inspire further experimental studies across diverse scenarios, bridging biological physics and engineering applications.

\ \\
\noindent{\bf Acknowledgments:} SM acknowledges financial support from IISER Mohali through a Senior Research Fellowship. PD acknowledges financial support from SERB, India through a start-up research grant (SRG/2022/000105). 

\ \\
\noindent{\bf Conflict of interest:} The authors have no conflicts to disclose.


\begin{thebibliography}{99}
\bibitem{S10} S. Ramaswamy, Annu. Rev. Condens. Matter. Phys. \textbf{1}, 323 (2010).

\bibitem{MJSTJR13} M. C. Marchetti, J. F. Joanny, S. Ramaswamy, T. B. Liverpool, J. Prost, M. Rao, and R. A. Simha, Rev. Mod. Phys. \textbf{85}, 1143 (2013).

\bibitem{J24}J. Toner, \textit{The Physics of Flocking} (Cambridge University Press, Cambridge, 2024).

\bibitem{CPDLG20} C. B. Caporusso, P. Digregorio, D. Levis, L. F. Cugliandolo, and  G. Gonnella, Phys. Rev. Lett. \textbf{125}, 178004 (2020).

\bibitem{JGMJC22} J. Tailleur, G. Gompper, M. C. Marchetti, J. M. Yeomans, and  C. Salomon, \textit{Active Matter and Nonequilibrium Statistical Physics, Lecture Notes of the Les Houches Summer School} \textbf{112}, (Oxford University Press, Oxford, 2022).

\bibitem{KGS23} \textit{Out-of-equilibrium Soft Matter}, edited by C. Kurzthaler, L. Gentile and H. A. Stone (The Royal Society of Chemistry, United Kingdom, 2023).

\bibitem{CRHCG16} C. Bechinger, R. Di Leonardo, H. L\"{o}wen, C. Reichhardt, G. Volpe, and G. Volpe, Rev. Mod. Phys. \textbf{88}, 045006 (2016).

\bibitem{AKTP21} A. K. Omar, K. Klymko, T. Grandpre, and P. L. Geissler, Phys. Rev. Lett. \textbf{126}, 188002 (2021).

\bibitem{YJRB23} Y. Duan, J. Agudo-Canalejo, R. Golestanian, and B. Mahult, Phys. Rev. Lett. \textbf{131}, 148301 (2023).

\bibitem{AJAYPJ23} A. Dinelli, J. \"{O}Byrne, A. Curatolo, Y. Zao, P. Sollich, and J. Tailleur, Nat. Commun. \textbf{14}, 7035 (2023).

\bibitem{SP24} S. Mondal and P. Das, J. Chem. Phys. \textbf{161}, 134902 (2024).

\bibitem{PMWBL12} P. Romanczuk, M. B\"{a}r, W. Ebeling, B. Lindner, and L. Schimansky-Geier, Eur. Phys. J. Special Topics. \textbf{202}, 1 (2012).

\bibitem{AH12} A. Patotsky and H. Stark, Europhys. Lett. \textbf{98}, 50004 (2012).

\bibitem{USAG18} U. Basu, S. N. Majumdar, A. Rosso, and G. Schehr, Phys. Rev. E, \textbf{98}, 062121 (2018).

\bibitem{IGFGC12} I. Buttinoni, G. Volpe, F. Kümmel, G. Volpe, and C. Bechinger, J. Phys.: Condens. Matter \textbf{24} 284129 (2012).

\bibitem{IJFHCT13} I. Buttinoni, J. Bialk\'{e}, F. K\"{u}mmel, H. L\"{o}wen, C. Bechinger, and  T. Speck, Phys. Rev. Lett. \textbf{110}, 238301 (2013).

\bibitem{AJRMYMJ13} A. P. Solon, J. Stenhammar, R. Wittkowski, M. Kardar, Y. Kafri, M. E. Cates and J. Tailleur, Phys. Rev. Lett. \textbf{110}, 238301 (2013).

\bibitem{JRG15}J. Elgeti, R. G. Winkler and G. Gompper, Rep. Prog. Phys. \textbf{78}, 056601 (2015).

\bibitem{AH23} A. Z\"{o}ttl and H. Stark, Annu. Rev. Condens. Matter Phys, \textbf{14}, 109 (2023).

\bibitem{GE36} G. E. Uhlenbeck and L. S. Ornstein, Phys. Rev. \textbf{36}, 823 (1930).

\bibitem{YM12} Y. Fily and M. C. Marchetti, Phys. Rev. Lett. \textbf{108}, 235702 (2012).

\bibitem{ECMJPF16} \'{E}. Fodor, C. Nardini, M. E. Cates, J. Tailleur, P. Visco, and F. v. Wijland, Phys. Rev. Lett. \textbf{117}, 038103 (2016).

\bibitem{MJ15} M. E. Cates and J. Tailleur,  Annu. Rev. Condens. Matter, \textbf{6}, 219 (2015).

\bibitem{LUAA19}L. Caprini, U. M. B. Marconi, A. Puglisi, and A. Vulpiani, J. Stat. Mech. 053203 (2019).

\bibitem{DJMCJF21} D. Martin, J. O'Byrne, M. E. Cates, \'{E}. Fodor, C. Nardini, J. Tailleur, and F. van Wijland, Phys. Rev. E \textbf{103}, 032607 (2021).

\bibitem{ERM22} \'{E}. Fodor, R. L. Jack and M. E. Cates, Annu. Rev. Condens. Matter Phys. \textbf{13},  215 (2022).

\bibitem{LDACH24} L. Caprini, D. Breoni, A. Ldov, C. Scholz, and H. L\"{o}wen, Commun. Phys. \textbf{7}, 343 (2024).

\bibitem{CMNALD14} C. Maggi, M. Paoluzzi, N. Pellicciotta, A. Lepore, L. Angelani, and R. Di. Leonardo, Phys. Rev. Lett. \textbf{113}, 238303 (2014).

\bibitem{CMLR17} C. Maggi, M. Paoluzzi, L. Angelani, and R. Di Leonardo, Sci. Rep. \textbf{7}, 17588  (2017).

\bibitem{TPJ15} T. F. F. Farage, P. Krinninger and J. M. Brader, Phys. Rev. E \textbf{91}, 042310 (2015).

\bibitem{ARJ17} A. Sharma, R.  Wittmann and J. M. Brader, Phys. Rev. E \textbf{95}, 012115 (2017).

\bibitem{H20} H. L\"{o}wen, J. Chem. Phys. \textbf{152}, 040901 (2020).

\bibitem{AALITASL0EM14} A. Attanasi, A. Cavagna, L. D. Castello, I. Giardina, T. S. Grigera, A. Jeli\'{c}, S. Melillo, L. Parisi, O. Pohl, E. Shen, and M. Viale, Nat. Phys. \textbf{10}, 691 (2014).

\bibitem{JRUMTHV20} J. Jhawar, R. G. Morris, U. R. Amith-Kumar, M. D. Raj, T. Rogers, H. Rajendran, and V. Guttal, Nat. Phys. \textbf{16}, 488 (2020).

\bibitem{MG13} M. Mijalkov and G. Volpe, Soft Matter. \textbf{9}, 6376 (2013).

\bibitem{MFJG18} M. Leyman, F. Ogemark, J. Wehr, and G. Volpe, Phys. Rev. E \textbf{98}, 052606 (2018).

\bibitem{0V19} O. Dauchot and V. D\'{e}mery, Phys. Rev. Lett. \textbf{122}, 068002 (2019).

\bibitem{GRH22} G. H. P. Nguyen, R. Wittmann and H. L\"{o}wen, J. Phys.: Condens. Matter.\textbf{34}, 035101 (2022).

\bibitem{LASHR22} L. Caprini, A. R. Sprenger, H. L\"{o}wen, and R. Wittmann, J. Chem. Phys. \textbf{156}, 071102 (2022).

\bibitem{LRH24} L. Caprini, R. K. Gupta and H. L\"{o}wen, Phys. Chem. Chem. Phys. \textbf{24}, 24910 (2022).

\bibitem{C85} C. W. Gardiner, \textit{Handbook of Stochastic Methods} (Berlin: Springer,1985) 

\bibitem{H96} H. Risken, \textit{The Fokker-Planck Equation}(Springer, Berlin, 1996).

\bibitem{V08} V. Balakrishnan, \textit{Elements of Nonequilibrium Statistical Mechanics} (CRC Press, 2008).

\bibitem{LVINCI} L. Da Vinci, \textit{Static measurements of sliding and rolling friction}, Codex Arundel, folios 40v, 41r, British Library.

\bibitem{C21} C. A. De Coulomb, \textit{Th\'{e}orie des machines simples,en ayant \'{e}gard au frottement de leurs parties et \'{a} la roideur des cordages} (reprinted by Bachelier, Paris, 1821).

\bibitem{P05} P. G. de Gennes, J. Stat. Phys. \textbf{119}, 953 (2005).

\bibitem{H05} H. Hayakawa, Physica D \textbf{205}, 48 (2005).

\bibitem{AN11} A. M. Menzel and N. Goldenfeld, Phys. Rev. E \textbf{84}, 011122 (2011).

\bibitem{PSM17} P. Das, S. Puri and M. Schwartz,  Eur. Phys. J. E \textbf{40}, 1 (2017).

\bibitem{RDAL98} R. I. Leine, D. H. van Campen, A. de Kraker, and L. van den Steen, Nonlinear Dyn. \textbf{16}, 41 (1998).

\bibitem{YM98} Y. Murayama and M. Sano, J. Phys. Soc. Jpn. \textbf{67}, 1826 (1998).

\bibitem{PSM16} P. Das, S. Puri and M. Schwartz, Phys. Rev. E \textbf{94}, 032907 (2016).

\bibitem{PSM18} P. Das, S. Puri and M. Schwartz,  Granular Matter. \textbf{20}, 15 (2018).

\bibitem{PP87} P. Jung and P. H\"{a}nggi, Phys. Rev. A \textbf{35}, 4464 (1987).

\bibitem{PP95} P. H\"{a}nggi and P. Jung, Adv. Chem. Phys. \textbf{89}, 239 (1995).

\bibitem{PW17} P. M. Geffert and W. Just, Phys. Rev. E \textbf{95}, 062111 (2017).

\bibitem{ALACH24} A. P. Antonov, L. Caprini, A. Ldov, C. Scholz, and H. L\"{o}wen, Phys. Rev. Lett. \textbf{133}, 198301 (2024).

\bibitem{SAS07} S. Andersson, A. S\"{o}derberg and S. Bj\"{o}rklund, Tribol. Intl. \textbf{40}, 580 (2007).

\bibitem{EVPP16} E. Pennestrì, V. Rossi, P. Salvini, and P. P. Valentini, Nonlinear Dyn. \textbf{83}, 1785 (2016).

\bibitem{CSA21} C. Wang, S. Tawfick and A. F. Vakakis, Phys. Rev. E \textbf{104}, 044906 (2021).

\bibitem{J09} J. F. Morris, Rheol. Acta \textbf{48}, 909 (2009).

\bibitem{TA23} T. Lequy and A. M. Menzel, Phys. Rev. E \textbf{108}, 064606 (2023).

\bibitem{DASU24} D. Dutta, A. Kundu, S. Sabhapandit, and U. Basu, Phys. Rev. E \textbf{110}, 044107 (2024).

\bibitem{DAU24} D. Dutta, A. Kundu and U. Basu, arXiv:2411.19186 (2024).

\bibitem{WR83} W. Horsthemke  and R. Lefever, \textit{ Noise-Induced Transitions: Theory and Applications in Physics, Chemistry and Biology} (Berlin: Springer,1983).

\bibitem{J84} J. M. Sancho, J. Math. Phys. \textbf{25}, 354 (1984).

\bibitem{I06} I. Bena, Int. J. Mod. Phys. \textbf{20}, 2825 (2006).

\bibitem{B10} B. Lindner, New J. Phys. \textbf{12}, 063026 (2010).

\bibitem{CE06} C. Kim and E. K. Lee,  Phys. Rev. E \textbf{73}, 026101 (2006).

\bibitem{zw19} X. Zhu and D. Wu,  Physica  A \textbf{514}, 259 (2019).

\end{thebibliography}
\end{document}